# Innovative Approaches for efficiently Warehousing Complex Data from the Web

**Fadila Bentayeb\*, Nora Maïz\*, Hadj Mahboubi\*\*, Cécile Favre\*, Sabine Loudcher\*, Nouria Harbi\*, Omar Boussaïd\* and Jérôme Darmont\***

*\* University of Lyon (ERIC-Lyon 2), France*
*first-name.last-name@univ-lyon2.fr*
*\*\*CEMAGREF Centre Clermont-Ferrand, France*
*hadj.mahboubi@cemagref.fr*

## ABSTRACT

Research in data warehousing and OLAP has produced important technologies for the design, management and use of information systems for decision support. With the development of Internet, the availability of various types of data has increased. Thus, users require applications to help them obtaining knowledge from the Web. One possible solution to facilitate this task is to extract information from the Web, transform and load it to a Web Warehouse, which provides uniform access methods for automatic processing of the data. In this chapter, we present three innovative researches recently introduced to extend the capabilities of decision support systems, namely (1) the use of XML as a logical and physical model for complex data warehouses, (2) associating data mining to OLAP to allow elaborated analysis tasks for complex data and (3) schema evolution in complex data warehouses for personalized analyses. Our contributions cover the main phases of the data warehouse design process: data integration and modeling and user driven-OLAP analysis.

## INTRODUCTION

Traditional databases aim at data management, i.e., they help organize, structure and query data. They are transaction processing-oriented and are often qualified as production databases. In opposition, data warehouses have a very different vocation: analyzing data (Kimball & Ross, 2002; Inmon, 2005) by exploiting specific models (star, snowflake and constellation schemas). They are termed as On-Line Analytical Processing (OLAP) databases. Data are then organized around indicators called measures, and analysis axes called dimensions. Dimension attributes either form a hierarchy or are just descriptive. Dimension hierarchies allow for obtaining views of data at different granularities, i.e., summarized or detailed through roll-up and drill-down operations, respectively.

Research in data warehousing and OLAP has produced important technologies for the design, management and use of information systems for decision support. To achieve the value of a data warehouse, incoming data must be transformed into an analysis-ready format. In the case of numerical data, data warehousing systems often provide tools to assist in this process. Unfortunately, standard tools are inadequate for producing relevant analysis when data are complex. Indeed, with the development of Internet, the availability of various types of data (Web data, multimedia data, biomedical data, etc.) has increased. Thus, users require applications to help them obtaining knowledge from the Web. For example, in the context of e-commerce, analyzing the behavior of a customer, a product, or a company consists of monitoring one or several activities (commercial or medical pursuits, patents deposits, etc.). The Web then becomes a real data source with which decision support applications should deal.

Furthermore, many Business Intelligence (BI) applications necessitate external data sources. For instance, performing competitive monitoring for a given company requires the analysis of data available only from its competitors. In this context, the Web is a tremendous source of data, and may be considered as a farming system.

However, the specific characteristics of Web data make it difficult to create such applications. One possible solution to facilitate this task is to extract information from the Web, transform and load it to a Web Warehouse, which provides uniform access methods for automatic processing of the data. Web Warehousing extends the life-time of Web contents and its reuse by different applications across time.

Moreover, the special nature of complex data poses different and new requirements to data warehousing technologies, over those posed by conventional data warehouse applications. In this case, the data warehousing process should be adapted in response to evolving complex data and information requirements. Tools must be developed to provide the needed analysis. Therefore, the issue that may arise "Can we OLAP complex data?" To address this issue, we need a new generation of data warehousing models that can organize complex data in a multidimensional way and new OLAP operators that can analyze them.

The XML formalism has emerged as a dominant W3C standard for describing and exchanging semistructured data among heterogeneous data sources. Its self-describing hierarchical structure enables a manipulative power to accommodate complex, disconnected and heterogeneous data. It allows describing the structure of a document and constraining its contents. With its vocation for semistructured data exchange, the XML language offers great flexibility for representing heterogeneous data, and great possibilities for structuring, modeling and storing them.

Furthermore, a data cube structure can provide a suitable context for applying data mining methods. More generally, the association of OLAP and data mining allows elaborated analysis tasks exceeding the simple exploration of a data cube. The aim is to take advantage of OLAP, as well as data mining techniques, and to integrate them into the same analysis framework in order to provide an enhanced analysis process for complex data by extending OLAP analysis capabilities.

In this chapter, we present three innovative researches recently introduced to extend the capabilities of decision support systems, namely (1) the use of XML as a logical and physical model for complex data warehouses, (2) the association of data mining with OLAP to allow elaborated analysis tasks for complex data and (3) the schema evolution in complex data warehouses for personalized analyses. Our contributions cover the main phases of the data warehouse design process: data integration and modeling and user driven-OLAP analysis.

1.      Our first contribution consists of using XML to model complex data warehouses (Boussaïd et al., 2006; Boussaïd et al., 2008). XML is suitable for structuring complex data coming from different Web sources and bearing heterogeneous formats (Mahboubi, 2009). XML indeed embeds both data and schema, either implicitly or explicitly through schema definitions (DTD or XML Schema). This type of metadata suits data warehouses very well. Moreover, XML query languages such XQuery help formulate analytical queries that would be difficult to express in a relational, SQL-based system (Beyer et al., 2005).

2.      Our second contribution deals with complex data analyzing based on data mining technologies. Classical OLAP tools are indeed ill-adapted to analyze Web data directly. OLAP facts representing complex objects need appropriate tools and new ways to be analyzed. Combining data mining methods with OLAP tools is an interesting solution to enrich OLAP capabilities and to analyze complex data from the Web. We propose to extend OLAP to describe, cluster and explain of complex data (Ben Messaoud et al., 2006, 2007a, 2007b).

3.      Our third contribution studies the problem of specifying changes in multidimensional databases. These changes may be motivated by evolutions of user requirements, as well as changes in Web sources (Hurtado et al., 1999b; Espil & Vaisman, 2001). The rule-based multidimensional model we provide supports both data and structure changes. The approach consists of creating new hierarchy levels in OLAP dimensions for data warehouse schema evolution according to relevant personalized analysis needs (Bentayeb et al., 2008; Favre et al., 2007).

Throughout this chapter, we use a running example to illustrate our contributions. It is extracted from a project we jointly carried out with linguist colleagues. This project (named CLAPI) dealt with the on-line integration, storage, management and analysis of spoken language interaction corpora (Aouiche et al., 2003). Funded by the French Ministry of Higher Education and Research, it provides Web access[1] to the CLAPI corpus database through user-friendly query and quantitative analysis tools (a new feature in spoken language research) that allow researchers in linguistics to elaborate hypotheses and validate results on a significant volume of data.

A conceptual schema of our excerpt from CLAPI is provided in Figure 1. A corpus is made of audio and/or video recordings of real-life interactions (e.g., classrooms, divorce conciliations, conflicts in queues…). Each speaker in a recording is identified (with a pseudo; the database is anonymous) and may appear in several interactions. To be exploited by linguists, recordings are textually transcribed. These transcriptions are actually structured in XML and feature both tokens (i.e., oral forms of words, such as "h'llo" for "hello") and interactional phenomena (e.g., pauses, laughs, speech overlaps…). Finally, scientific studies produced by researchers in linguistics are attached to the corresponding recordings, which may be several when the study is transversal. All this information is available on the Web and stored with the help of standard Web technologies such as XML and its derivatives (e.g., RDF for metadata).

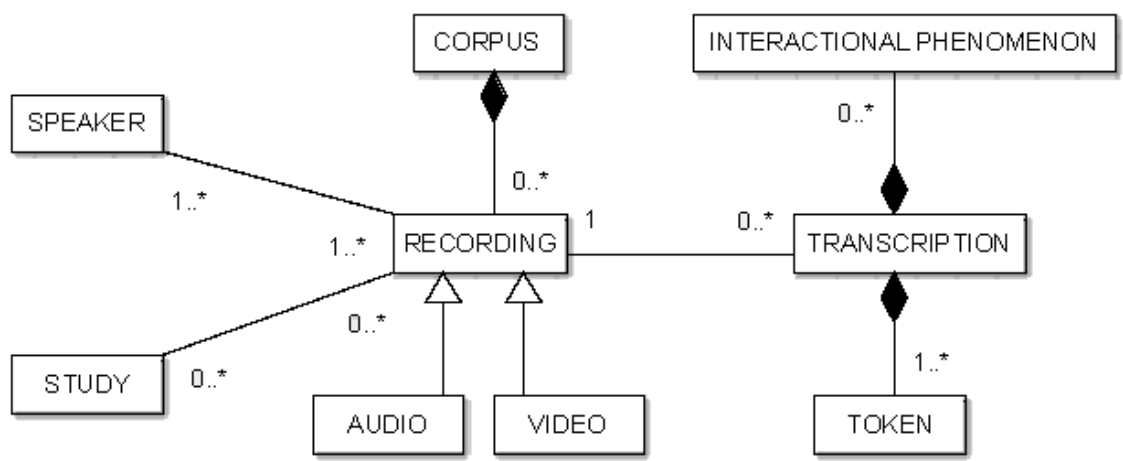

*Figure 1.CLAPI excerpt UML class diagram*

In this context, OLAP is able to provide linguist users statistical information regarding analysis dimensions such as speakers, tokens, phenomena or time elapsed within a single interaction. It also helps compare several similar interactions (e.g., classrooms, conflicting situations…). For instance, simple analysis scenarios include:

- count the frequency of a given token over a determined period of time;
- count the frequency and evolution of speech overlaps over a determined period of time;
- number identified phenomena (such as laughs) over a determined period of time;
- compute the total, average, maximum and/or minimum pause time spent by speakers over a determined period of time.

Answering such queries requires fetching and integrating source data from the Web, dealing with data complexity and building suitable (XML) data cubes onto which suitable OLAP operations must be applied.

The remainder of this chapter is organized as follows. First, we discuss the related work regarding complex Web data warehousing in Section 2. Then, we detail our three contributions in Sections 3, 4, and 5. We finally conclude this chapter and provide future research directions in Section 6.

## GENERAL LITERATURE REVIEW

### Complex data integration

In data warehousing, the prime objective of storing data is to facilitate the decision process. To achieve the value of a data warehouse, input data must be transformed into an analysis-ready format. In the case of numerical data, data warehousing systems often provide tools to assist in this process. Unfortunately, standard tools are inadequate for producing a relevant analysis axis when data are complex. In such cases, the data warehousing process should be adapted in response to evolving data and information requirements. In a data warehousing process, the data integration phase is crucial. Data integration is a hard task that involves reconciliation at various levels: data models, data schema, data instances, and semantics.

Two main and opposing approaches are used to perform data integration over heterogeneous data sources. In the mediator-based approach (Rousset, 2002), data remain located at their original sources. User queries are executed through a mediator-wrapper system (Goasdoué et al., 2000). A mediator reformulates queries according to the content of the various accessible data sources, while the wrapper extracts the selected data from the target source. The major advantage of this approach especially in a Web context is its flexibility, since mediators are able to reformulate and/or approximate queries to better satisfy the user. However, when data sources are updated, modified data are lost, which is not pertinent in a decision-support context where data historicity is important.

In opposition, in the data warehouse approach, selected data from various sources are centralized in a new multidimensional database, the data warehouse (Inmon, 2005; Kimball, 2002). In a data warehouse context, data integration corresponds to the Extract, Transform, and Load (ETL) process that accesses, cleans and transforms heterogeneous data before they are loaded into the data warehouse. This approach supports data dating and is tailored for analysis. Some studies combine the mediation-based integration and data warehousing (Rousset, 2002). Other authors propose a logic description framework or a language (e.g., CARIN) for information integration (Goasdoué et al., 2000; Calvanese et al., 1998).

The special nature of complex data poses different and new requirements to data warehousing technologies, over those posed by conventional data warehousing applications. Hence, to integrate complex data sources, we need more than a tool for organizing data into a common syntax. The integration of complex data in a database is an issue that has been little studied. Jensen et al. (2001) propose a general system architecture for integrating XML and relational data sources at the conceptual level in a Web-based OLAP database. A "UML snowflake diagram" is built by choosing the desired UML classes from any source. The process is deployed through a graphical interface that deals only with UML classes and makes data sources transparent to the designer. Other approaches for an easier integration of complex data semantics have also been published. A framework for video content understanding, consisting of an expert system that uses a rule-based engine, domain knowledge, visual detectors and metadata to enhance video detection results and presenting the semi-automatic construction of multimedia ontologies, is presented in (Jaimes et al., 2003). Semantic indexing techniques for complex data also exist (Stoffel et al., 1997). These techniques are based on domain knowledge made available in the form of ontologies. This research overcomes the difficulty of efficiently integrating semantic knowledge stored as ontologies in ways that support the efficient indexing of large databases. Finally, Baumgartner et al. (2005) describe how public information can be extracted automatically from Web sites, transformed into structured data formats, and used for data analysis in Business Intelligence systems. Authors design an architecture, called Lixto, for processing Web data in an efficient manner. They also exploit the XML formalism to manage extracted data and make them exploitable into an SAP business information warehouse.

To process complex data, it is also important to consider their metadata. To take this information into account, it is necessary to use appropriate management tools. An integrative and uniform model for metadata management in data warehousing environments uses a uniform representation approach based on UML to integrate technical and semantic metadata and their interdependencies (Stohr et al., 2002). Standards for describing resources that we consider as complex data already exist. The Resource Description Framework (Lassila & Swick, 1999) is a W3C-approved standard that uses metadata to describe Web contents. It is possible to describe all kind of resources using RDF. Another important standard is MPEG-7 (Manjunath et al., 2002). Its most important goal is to provide a set of methods and tools for the different aspects of multimedia content description. MPEG-7 focuses on the standardization of a common interface for describing multimedia materials (representing information about contents and metadata).

## Data warehouse modeling

Multidimensional modeling aims at representing data according to user requirements. It represents information to be analyzed as a set of points in a space with several dimensions that are analysis axes.

Currently, although no model is recognized as standard, three concepts are admitted by the data warehousing community as the bases of multidimensional modeling: the concept of "data cube" and those of "dimension" and "fact". The first suggested models are based on structures (cubes) that do not represent all the concepts expressed by users, but the characteristics of their implementation. To define a multidimensional model that represents the concepts of the real world independently from any physical aspect, other concepts must be added. Multidimensional analyses require concepts that are close to the vision of decision makers, and to the semantics of decision. The introduced concepts are: hierarchies, various types of star, snowflake and constellation schemas (Kimball 2002), parameters, and weak attributes (Golfarelli et al., 1998).

However, dimensional modeling must be adapted to take into account the specificities of complex data. Some proposals regard multidimensional modeling by using XML as a base language for describing data warehouses. Krill affirms that vendors such as Microsoft, IBM, and Oracle largely employ XML in their database systems for interoperability between data warehouses and tool repositories (Krill, 1998). Nevertheless, we distinguish two separate approaches in this field.

The first approach focuses on the physical storage of XML documents in data warehouses. XML populates warehouses since it is considered an efficient technology to support data within well suited structures for interoperability and information exchange. Baril and Bellahsène introduce the View Model (Baril, & Bellahsène, 2003), which is a method capable of querying XML databases. A data model is defined for each view to organize semi-structured data. An XML warehouse, named DAWAX (DAta WArehouse for XML), based on the View Model is also proposed. Hümmer et al. (2003) propose an approach, named XCube, to model classical data cubes with XML. Nevertheless, this approach focuses on the exchange and the transportation of classical data cubes over networks rather than on multidimensional modeling with XML.

The second approach aims at using XML to design data warehouses according to classical multidimensional models such as star and snowflake schemas. XML-star schema (Pokorný, 2001) uses Document Type Definitions (DTDs) to explicit dimension hierarchies. A dimension is modeled as a sequence of DTDs that are logically associated similarly as referential integrity does in relational databases. Golfarelli et al. (1998) introduce a Dimensional Fact Model represented via Attribute Trees. They also use XML Schemas to express multidimensional models by including relationships with sub-elements (Golfarelli et al. 2001).

However, Trujillo et al. (2004) think that this approach focuses on the presentation of multidimensional XML rather than on the presentation of the structure of the Multidimensional Conceptual Modeling (MCM) itself. They claim that an Object Oriented (OO) standard model is rather needed to cope with all multidimensional modeling properties at both structural and dynamic levels. Trujillo et al. (2004) provide a DTD model from which valid XML documents are generated to represent multidimensional models at a conceptual level. Nassis et al. (2004) propose a similar approach where OO is used to develop a conceptual model for XML Document Warehouses (XDW). An XML repository, called xFACT, is built by integrating OO concepts with XML Schemas. Nassis et al. (2004) also define Virtual dimensions by using XML and UML package diagrams in order to help the construction of hierarchical conceptual views. The X-Warehousing process is entirely based on XML: it designs warehouses with XML Schemas at a logical level, and then populates them with valid XML documents at a physical level (Boussaïd et al., 2006). Further, since it uses XML, our approach can also be considered a real solution for warehousing heterogeneous and complex data in order to prepare them for future OLAP analysis.

## OLAP

Online Analytical Processing  (OLAP) is an efficient technique for explorative data analysis. This trend is obvious, given the popularity of many OLAP systems (Codd, 1993), such as Essbase (Arbor Software) or Express (Oracle Corporation) for instance. Based on a multidimensional conceptual view of the data, these systems are especially well-suited to data analysis, and their characteristics are significantly different from those of relational databases. The analysis process concerns basic or aggregated data containing relevant information. OLAP allows to model data in a dimensional way and to observe data from different perspectives. This approach consists of building data cubes (or hypercubes) on which OLAP operations are performed. A data cube is a set of facts described by measures to be observed along analysis axes (dimensions) (Kimball, 2002; Inmon, 2005; Chaudhuri & Dayal, 1997). A dimension may be expressed through several hierarchies. Many aggregation levels for measures can be achieved to obtain either summarized or detailed information using OLAP operators. Thus, hierarchies allow sophisticated analyses and data visualization in a multidimensional database (Jagadish et al., 1999).  However, the classical OLAP operators are only well-suited to numerical data, whereas, in recent years, more and more data sources beyond conventional alphanumerical data have come into being. This phenomenon especially appears when we wish to analyze and aggregate such objects as chemical compounds or protein networks (chem/bio-informatics), 2D/3D objects (spatial/geographic data and pattern recognition), circuits (computer-aided design, simulation), XML data (with loose schemas) and Web activities (human/computer networks and interactions). Such data are not only individual entities. Interacting relationships among them are also important and interesting. These relationships carry more semantics than entities themselves. Exploiting the semantics of complex data can then build decision-making information that is more relevant than OLAP classical approach allows.

In this context, several efforts have made done to define adequate complex data analysis. Some are spatially dedicated to define adequate operators for multimedia and spatial data analysis. For instance, Bimonte et al. (2007) propose a Web-based GIS-OLAP integrated solution supporting geographical dimensions and measures, and providing interactive and synchronized maps, pivot tables and diagrams displays in order to effectively support decision makers. Arigon et al. (2007) help user select the best representation of medical data according to various functional versions of the dimension numbers. Others studies aim at performing OLAP analyses over XML data (XOLAP) representing complex data. Such research work proposes to extend the XQuery with near OLAP capabilities such as advanced grouping and aggregation features. For instance, Hachicha et al. (2008) propose to express a broad set of OLAP operators with the TAX XML algebra.
All these contributions have a common point. They are based on their specific areas to adapt existing tools to make them suitable for the analysis of their requirements.
However, the limitations of traditional OLAP remain. It is the use of other techniques, such as data mining or information retrieval, combined with on-line analysis, that help obtain new analytical capabilities to process complex data.

The most recent proposals are especially interested in the OLAP analysis of social networks, interactions, or the Web 2.0 (Sifer, 2005; Aouiche et al., 2009). For example, Sifer (2005) propose to explore Web logs using a specific representation rather than data cubes. OLAP operations are based on coordinated dimension views and correspond to selection operations. Aouiche et al. (2009) give more attention to analyzing tag-clouds. Such data are becoming usual in so-called Social Web applications. For this sake, the authors define a set of rules to formally recognize an OLAP application and define a set of adequate OLAP computation of tag clouds as top-k queries.

## XML DATA WAREHOUSING

Already standard on the Web, the XML language has also become a standard for representing business data (Beyer et al., 2005). It is particularly adapted for modeling complex data originating from heterogeneous sources, particularly from the Web. Moreover, XML query languages such as XQuery help formulate analytical queries that would be difficult to express in a relational system (Beyer et al., 2005). In consequence, there has been a clear trend toward XML warehousing for a couple of years. In this trend, we propose to represent complex data from the Web as XML documents. Then, a multidimensional model is designed to obtain an XML data warehouse. Finally, on-line analysis (OLAP) can take place.

## Related work: XML data warehousing and OLAP

Research in XML warehousing may be subdivided into three families. The first focuses on Web data integration for decision-support purposes. Web data sources are described by XML Schemas that are transformed into graphs used for selecting facts and creating a logical schema that validates the data warehouse (Golfarelli et al., 2001; Vrdoljak et al., 2003). In addition, the Xyleme system also supports query evaluation and change control (Xyleme, 2001). In these approaches, no particular warehouse model is proposed. All these proposals are reviewed by Pérez and al. (2008). Here, authors mainly focus on studies considering XML and its extensions as canonical formalisms for Web application interoperability: meta-data representation, data interchange and heterogeneous and distributed Web data analysis (OLAP). They survey works dealing with the integration of XML data from heterogeneous and distributed sources (data warehouses), XML multidimensional design (semi-automatic ways) and XML-compatible OLAP systems. Authors also discuss approaches addressing how OLAP and Information Retrieval (IR) can be combined to explore text-rich document collections and to analyze acts and documents together in so-called contextualized warehouses.

By contrast, the second family of XML warehousing approaches explicitly bases on classical warehouse logical models (star-like schemas) and supports end-user analytical tools. An XML warehouse is then composed of documents representing facts and dimensions (Pokorný, 2001; Hümmer et al., 2003; Park et al., 2005). A methodological effort has also been made to cover processes such as data cleaning, summarization, intermediating XML documents, updating/linking existing documents and creating fact tables (Rusu et al., 2005), or to represent user analysis needs and match them with source data (Boussaïd et al., 2006). All these approaches more or less converge toward a unified XML warehouse model. They mostly differ in the way dimensions are handled and the number of XML documents used to store facts and dimensions.

Finally, the third XML warehouse family relates to document warehouses. Here, an XML data warehouse is a collection of either materialized or virtual XML views, which provide a mediated schema that constitutes a uniform query interface (Baril and Bellahsène, 2003; Rajugan et al., 2005; Nassis et al., 2005; Zhang et al., 2005).

Though XML data warehousing issues are quite broadly addressed, fewer authors actually push through the whole decision-support process and address the multidimensional analysis of XML data, which is termed XML-OLAP or XOLAP. This is mainly achieved by extending existing languages such as Microsoft MDX or XQuery with XML-specific operators (Park et al., 2005) or grouping, numbering, aggregation and cube operators (Beyer et al., 2005; Wang et al., 2005; Wiwatwattana et al. 2007), respectively. In the same frame of mind, some proposals extend existing OLAP clients to incorporate dimensions and measures extracted from external XML sources (such as Web pages) into a data cube (Pedersen et al., 2006). Such extensions are exploited as ordinary dimensions and measures, and allow handling unexpected or short time data requirements. They involve novel multi-granular data models and query languages that formalize and extend the existing system (Pedersen et al., 2006).

## XML warehousing and analysis methodology

Though feeding data warehouses with XML documents from the Web is getting increasingly common, methodological issues arise. The multidimensional organization of data warehouses is indeed quite different from the semi-structured organization of XML documents. Their architecture is subject-oriented, integrated, consistent, and data are regularly refreshed to represent temporal evolutions. An XML formalism can definitely be used to describe the various elements of a multidimensional model (Boussaïd et al., 2006), but XML can only be considered as a logical and physical description tool for future analysis tasks. The reference conceptual model remains the star schema and its derivatives.

Hence, we propose an XML multidimensional (and thus analysis-oriented) model to derive a physical organization of XML documents. To support this choice, we propose a modeling process that achieves complex data integration. We first design a conceptual UML model for a complex object. This UML model is then directly translated into an XML Schema, which we view as a logical model. At the physical level, XML documents that are valid against this logical model may be mapped into a relational, object-relational or XML-native database. After representing complex data as XML documents, we physically integrate them into an Operational Data Store (ODS), which is a buffer ahead of the actual warehouse.

At this stage, it is already possible to mine the stored XML documents directly, e.g., with XML structure mining techniques. In addition, to further analyze these documents' contents efficiently, it is interesting to warehouse them, i.e., devise a multidimensional model that allows OLAP analyses. However, classical OLAP operators cannot handle XML's complexity and XOLAP operators are still few. Thus, we also propose a solution at this level.

## XML data warehouse model

Existing XML warehouse models mostly differ at the logical and physical levels in the number of XML documents used to store facts and dimensions. A performance evaluation study of these alternatives shows that representing facts in one single XML document and each dimension in one XML document allows the best performance in a snowflake schema (Boukraa et al., 2006). Moreover, this representation also allows to model constellation schemas without duplicating dimension information. Several fact documents can indeed share the same dimensions. Furthermore, since each dimension and its hierarchy levels are stored in one XML document, dimension updates are more easily and efficiently performed than if dimensions were either embedded with the facts or all stored in one single document.

Hence, we propose to adopt this architecture to represent XML data warehouses (Mahboubi et al., 2009). It is actually the translation of a classical constellation schema into XML. More precisely, as XCube (Hümmer et al., 2003), our reference data warehouse model is composed of the three types of XML documents: dw-model.xml represents warehouse metadata (schema, including dimension hierarchies); a set of facts$_f$.xml documents help store facts, i.e., dimension references and measure values; and a set of dimension$_d$.xml documents help store dimension member values.

Figure 2 represents dw-model.xml's graph structure. Its root node, DW-model, is composed of dimension and FactDoc nodes. A dimension node defines one dimension, its possible hierarchical levels (Level elements) and attributes (including their types), as well as the path to the corresponding dimensiond.xml document. A FactDoc element defines a fact, i.e., its measures, internal references to the corresponding dimensions, and the path to the corresponding factsf.xml document. Figure 3(a) represents the factsf.xml documents' graph structure. It is composed of fact nodes defining measures and dimension references. The document root node, FactDoc, is composed of fact subelements, each of whose instantiates a fact, i.e., measure values and dimension references. These identifier-based references support the fact-to-dimension relationship. Finally, Figure 3(b) represents the dimensiond.xml documents' graph structure. Its root node, dimension, is composed of Level nodes. Each defines a hierarchy level composed of instance nodes that in turn define the level's member attribute values. In addition, an instance element contains Roll-up and Drill-Down attributes that define the hierarchical relationship within the dimension.

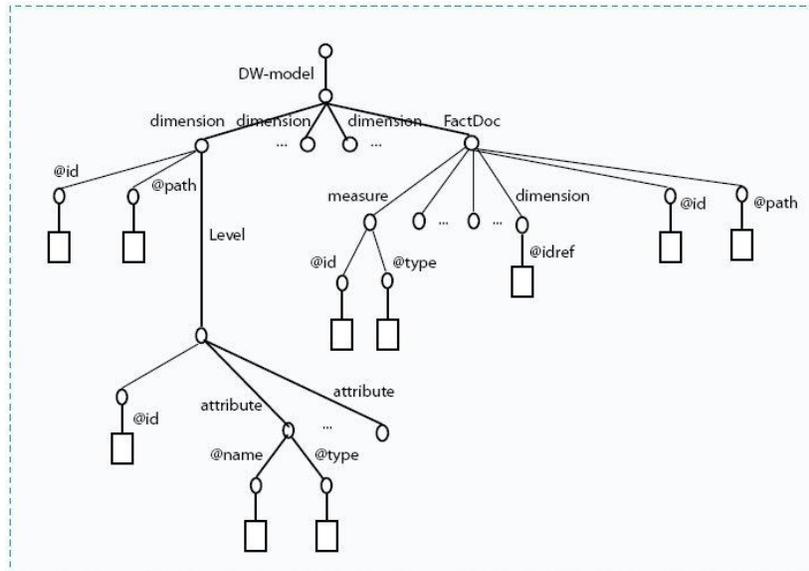

*Figure 2. dw-model.xml graph structure*

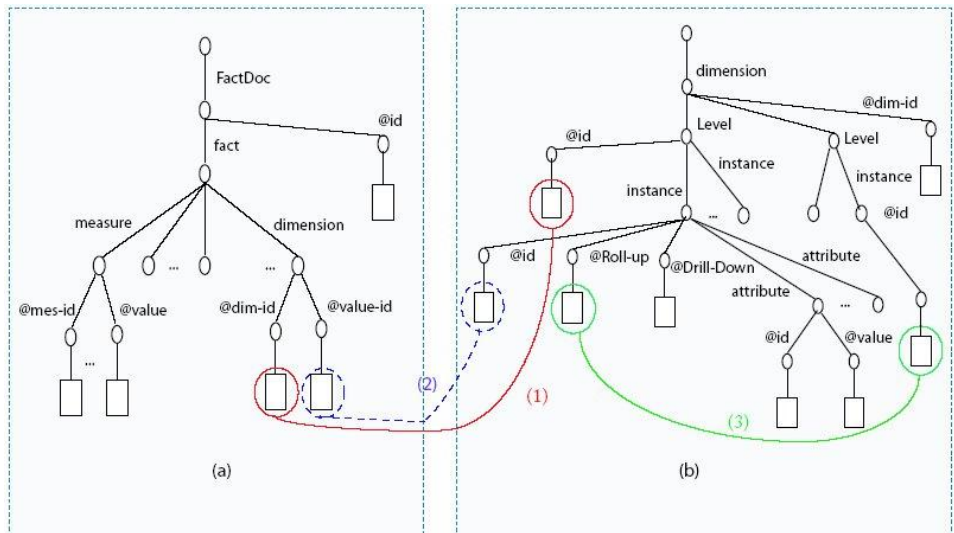

*Figure 3. factsf.xml (a) and dimensiond.xml (b) graph structures*

An example of instantiated dw-model.xml document is provided in Figure 4. It refers to a CLAPI-based study we made with our linguist colleagues to observe token (term in a transcription) frequencies, with respect to location in the transcription (begin, middle, end) and speaker sex. Due to space constraints, we cannot provide all dimensions and fact instances in this chapter, but the reader can extrapolate from Figure 4 the contents of dim-time.xml, dim-speaker.xml, dim-transcript.xml and facts.xml, respectively.

```
<?xml version="1.0" encoding="utf-8">
<DW-model>
    <dimension id="time-d" path="dim-time.xml">
        <Level id="location-in-transcription">
            <attribute name="location" type="string" />
        </Level>
    </dimension>
    <dimension id="speaker-d" path="dim-speaker.xml">
        <Level id="speaker">
            <attribute name="sex" type="boolean" />
        </Level>
    </dimension>
    <dimension id="transcription-d" path="dim-transcript.xml">
        <Level id="token">
            <attribute name="term" type="string" />
        </Level>
        <Level id="transcription">
            <attribute name="transcription-name" type="string" />
        </Level>
    </dimension>
    <FactDoc id="facts" path="facts.xml">
        <measure id="frequency" type="real" />
        <dimension idref="time-d" />
        <dimension idref="speaker-d" />
        <dimension idref="transcription-d" />
    </FactDoc>
</DW-model>
```

*Figure 4. Sample dw-model.xml document*

## Algebraic expression of XOLAP operators

This last decade's efforts for formalizing OLAP algebras have helped design a formal framework and well-identified operators. Existing OLAP operators, previously defined in either a relational or multidimensional context, are now being adapted to the data model of XML documents (i.e., graphs) and enriched with XML-specific operators. However, most existing approaches that aim at XOLAP do not fully satisfy these objectives. Some favor the translation of XML data cubes in relational, and query them with extensions of SQL. Others tend toward multidimensional solutions that exploit XML query languages such as XQuery. However, in terms of algebra, this work only proposes a fairly limited number of operators.

As Wiwatwattana et al. (2007), we aim at an XML-native solution that exploits XQuery. For this sake, we use the rich TAX Tree Algebra for XML (Jagadish et al., 2001) to support OLAP operators. We can express in TAX the main usual OLAP operators: cube, rotate, switch, roll-up, drill-down, slice, dice, pull and push (Hachicha et al., 2008). By doing so, we significantly expand the number of available XOLAP operators since, up to now, related papers only proposed at most three operators each (always including cube). We have also implemented these XOLAP operators into a software prototype that helps generate the corresponding XQuery code. This querying interface is currently coupled to TIMBER XML-native DBMS, but it is actually independent and could operate onto any other DBMS supporting XQuery.

In the next step of our work, we are taking inspiration from the principle of $X^3$ (Wiwatwattana et al., 2007) to enhance our XOLAP operators and truly make them XML-specific. For instance, we are currently working on performing roll-up and drill-down operations onto ragged hierarchies (Beyer et al., 2005).

# COMBINING OLAP AND DATA MINING FOR COMPLEX DATA ANALYSIS

We address in this section the issue of Web data analysis. Among decision-support technologies, OLAP offers techniques to visualize, summarize, and explore data. However, classical OLAP tools are unsuitable and unable to deal with complex data originating from heterogeneous sources, particularly from the Web. For example, when processing images, sounds, videos, texts or even XML documents, aggregating information with classical OLAP does not make any sense. We are not able to compute a sum or an average operation over such data. Hence we think that OLAP facts representing complex objects need appropriate tools and new ways to be analyzed. Besides, OLAP does not provide automatic tools to investigate interesting patterns from multidimensional data. In order to enrich its traditional capabilities and to analyze complex data from the Web, we propose to associate the OLAP technology with data mining techniques. Our general approach states that OLAP and data mining should be fully merged and considered an integral part of a unified analysis process. We propose three main approaches based on coupling OLAP and data mining. These approaches aim at extending OLAP to description, clustering, and explanation. In this part of the article, we briefly present these approaches: their objectives, their principles, and their results.

## Related work: coupling OLAP and data mining

The major difficulty when combining OLAP and data mining is that traditional data mining algorithms are mostly designed for tabular datasets organized in individual-variable form. Therefore, multidimensional data are not suited for these algorithms. Nevertheless, a lot of previous studies motivated and proved an interest for coupling OLAP with data mining methods. Ramakrishnan et al. (2007) discuss a class of new problems and techniques that show great promise for exploratory mining in cube spaces. We distinguish three major approaches in this field.

The first approach tries to extend the query language of decision support systems to achieve data mining tasks. The DBMiner system, proposed by Han (1998), summarizes this approach. Some extended OLAP operators feature data mining methods such as association, classification, prediction, clustering and sequencing. Han defines OLAP Mining as a new concept that integrates OLAP technology with data mining techniques and allows performing analyses on different portions and levels of abstraction in a data cube. He also introduces OLAM (On-Line Analytical Mining) as a process for extracting knowledge from multidimensional databases. He expects that, in the future, OLAM will be a natural addition to OLAP technology that enhances the power of multidimensional data analysis. Chen et al. (2000) discover behavior patterns by mining association rules about customers from transactional e-commerce data. They extend OLAP functions and use a distributed OLAP server with a data mining infrastructure. The resulting association rules are represented in particular cubes called Association Rule Cubes. Goil & Choudhary (1998) think that dimension hierarchies can be used to provide interesting information at multiple concept levels. Their approach summarizes information in a data cube, extends OLAP operators and mines association rules. Some other researches consist of integrating mining functions in the database system using SQL. Chaudhuri et al. (1999) propose a data mining system based on extending SQL and develop a client-server middleware that performs a decision tree classifier in MS SQL Server 7.0.

The second approach consists of adapting multidimensional data inside or outside the database system and applies classical data mining algorithms on the resulting datasets. This approach can be viewed with respect to two strategies. The first one consists of taking advantage from multidimensional database management systems (MDBMS) to help construct learning models. The second strategy transforms multidimensional data and makes them usable by data mining methods. For instance, Pinto et al. (2001) integrate multidimensional information in data sequences and apply on them frequent pattern discovery.

In order to apply decision trees on multidimensional data, Goil & Choudhary (1998) flatten data cubes and extract a contingency matrix for each dimension at each construction step of the tree. Chen et al. (2001) think that OLAP should be adopted as a pre-processing step in the knowledge discovery process. In the same context, Maedche et al. (2000) combine databases with classical data mining systems by using OLAP engine as interface to process telecommunication data. In this interface, OLAP tools create a target data set to generate new hypotheses by applying data mining methods. Tjioe & Taniar (2005) propose a method for mining association rules in data warehouses. Based on the multidimensional data organization, this method is capable of extracting associations from multiple dimensions at multiple levels of abstraction by focusing on measurements of summarized data. In order to do this, the authors propose to prepare multidimensional data for the mining process according to four algorithms: VAvg, HAvg, WMAvg, and ModusFilter. Fu (2005) proposes an algorithm, called CubeDT, for constructing decision tree classifiers based on data cubes. This algorithm works on statistic trees which are representations of multidimensional data especially suitable for the construction of decision trees.

The third approach is based on adapting data mining methods and applying them directly on multidimensional data. Palpanas (2000) thinks that adapting data mining algorithms is an interesting solution to provide elaborated analysis and precious knowledge. Parsaye (1997) claims that decision-support applications must consider data mining within multiple dimensions. He proposes a theoretical OLAP Data Mining System that integrates a multidimensional discovery engine in order to perform discovery along multiple dimensions. Sarawagi et al. (1998) propose to integrate a multidimensional regression module, called Discovery-driven, in OLAP servers. This module guides the user to detect relevant areas at various hierarchical levels of a cube. Imielinski et al. (2002) propose a generalized version of association rules called Cubegrades. The authors claim that association rules can be viewed as the change of an aggregate's measure due to a change in the cube's structure. Dong et al. (2001) enhance Cubegrades and introduce constrained gradient analysis. Their proposition focuses on extracting pairs of cube cells that are quite different in aggregates and similar in dimensions. Instead of dealing with the whole cube, constraints on significance, probability, and gradient are added to limit the search range. Finally, Chen et al. (2005) introduce *Prediction Cube*. In contrast to standard cubes, in which each cell value is computed by an aggregate function (e.g., SUM or AVG), each cell value in a prediction cube summarizes a predictive model trained on the data corresponding to that cell, and characterizes its decision behavior, or predictiveness.

This previous work has proved that associating data mining to OLAP is a promising way to allow the power of elaborated analysis tasks. This affirms that data mining methods are able to extend OLAP analyze. In addition to this work, we have proposed three main contributions to this field.

## OLAP aggregation by clustering

When users analyze complex data from the Web, they need more expressive aggregates than those created from the elementary computation of additive measures. We think that OLAP facts representing complex objects need appropriate tools and new ways of aggregation since to be analyzed. To summarize information about complex data, we should gather similar facts into a single group and separate dissimilar facts into different groups. In this case, it is necessary to consider an aggregation by computing both descriptors and measures. Instead of grouping facts only by computing their measures, we also take their descriptors into account to obtain aggregates expressing semantic similarities. Facts can be aggregated with a sum or average function but it is not interesting in the Web data context. However, it would be more interesting if facts seeming similar could be aggregated.

In order to do so, we couple OLAP with data mining to create a new type of online complex data aggregation. We have already proposed a new OLAP operator, called OpAC (Operator for Aggregation by Clustering), that combines OLAP with an automatic clustering technique (Ben Messaoud et al., 2006). We use the Agglomerative Hierarchical Clustering (AHC) as an aggregation strategy for complex data. We proved the interest of this new operator and its efficiency in creating semantic aggregates. More generally, the aggregates provided by OpAC give interesting knowledge about the analyzed domain. Our operator deals with all types of data by handling a data cube modeled by XML. For example, the new operator enables to note that some tokens like "Hello", "Hi" and "Good morning" form a significant aggregate since they have the same location in the transcription (they are at the beginning).

Furthermore, we also propose some evaluation criteria that help validate our operator. These criteria aim at assisting the user and helping him/her to choose the best partition of aggregates that fits with his/her analysis requirements.

## Multiple correspondence analysis to organize data cubes

OLAP provides the user with visual tools to summarize, explore and navigate into data cubes in order to detect interesting and relevant information. However, exploring a data cube is not always an easy task. Obviously, in large cubes containing sparse data, the whole analysis process becomes tedious and complex. In such a case, an intuitive exploration based on the user's experience does not quickly lead to efficient results. Current OLAP provides query-driven and visual tools to browse data cubes, but does not deeply assist the user and help him/her to investigate interesting patterns.

We propose to provide the user with an automatic assistance to identify interesting facts and arrange them in a suitable visual representation. We suggest an approach that allows the user to get relevant facts expressing relationships and displays them in an appropriate way that enhances the exploration process independently from cube size (Ben Messaoud et al., 2007b). Thus, we carry out a Multiple Correspondence Analysis (MCA) on a data cube as a preprocessing step. Basically, MCA is a powerful describing method even for huge volumes of data. It factors categorical variables and displays data in a factorial space constructed by an orthogonal system of axes that provides relevant views of data. These elements motivate us to exploit the results of MCA in order to better explore large data cubes by identifying and arranging interesting facts. The first constructed factorial axis summarizes the maximum of information contained in the cube. We focus on relevant OLAP facts associated with characteristic attributes (variables) given by the factorial axes. These facts are interesting since they reflect relationships and concentrate significant information. For a better visualization of these facts, we highlight them and arrange their attributes in the data space representation by using test-values. For example, consider the cube of Figure 5.

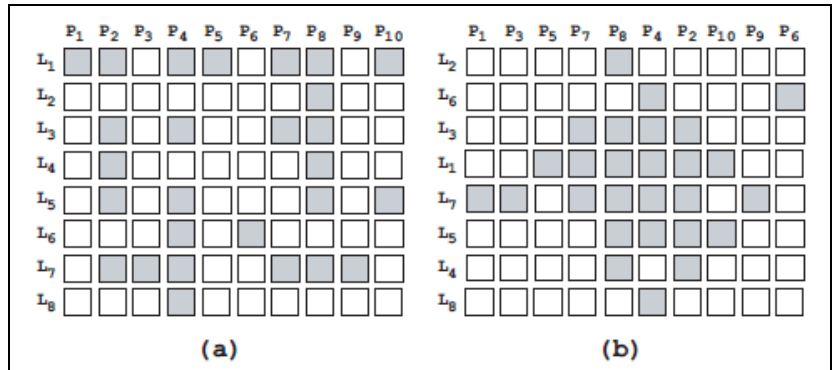

*Figure 5. Two representations of a cube*

On the one hand, representation 5(a) displays frequencies of tokens (P1, . . . , P10) crossed by locations in the transcription (L1, . . . , L8). In this representation, full cells (gray cells) are displayed randomly according to lexical order of tokens and of locations. On the other hand, Figure 5(b) contains the same information as Figure 5(a) but it displays a data representation visually easier to analyze. Figure 5(b) expresses important relationships by providing a visual representation that gathers full cells together and separates them from empty cells.

To evaluate the visual relevance of multidimensional data representations, we also propose a new criterion to measure the homogeneity of cell distribution in the representation space of a data cube. This criterion is based on geometric neighborhood of data cube cells, and also takes the similarity of cell measures into account and provides a scalar quantification for the homogeneity of a given data cube representation. It also allows evaluating the performance of our approach by comparing the quality of the initial data representation and the arranged one.

## Association rules for explanation in data cubes

OLAP techniques do not allow the identification of relationships, groupings or exceptions that could hold in a data cube. To that end, we propose to extend OLAP tools to explanation capabilities by mining association rules in multidimensional data. This new approach is capable of automatically explaining relationships and associations that could exist in data cubes. In the study we made with our linguist colleagues, the frequency of a token can be particularly high at the end of transcriptions. This frequency can be explained by an association between the token, the location in the transcription and the speaker sex. For example, this token is often use by women to finish a talk.

Frequent itemset, sequential itemset and association rule mining help achieve this goal.
We propose an on-line environment for mining association rules in data cubes (Ben Messaoud et al., 2007a). We use the concept of inter-dimensional meta-rule that allows users to guide the mining process and focus on a specific context from which rules can be extracted. Our framework also allows a redefinition of the support and confidence measures based on aggregate functions (SUM and COUNT) used as cube indicators (measures). Therefore, the computation of support and confidence according to the COUNT measure becomes a particular case in our proposal. In addition to support and confidence, we use two other descriptive criteria (Lift and Loevinger) to evaluate the interestingness of mined associations. These criteria are computed according to a sum-based aggregate measure in the data cube and reflect interestingness of associations in a more relevant way than what is offered by support and confidence. We develop our proposal according to a bottom-up algorithm for searching association rules. Our algorithm consists of an adaptation of the traditional Apriori algorithm to multidimensional data.

In addition, in order to focus on the discovered associations and validate them, we provide a visual representation based on the graphic semiology principles. Such a representation consists of a graphic encoding of frequent patterns and association rules in the same multidimensional space as the one associated with the mined data cube.

To improve decision support systems and to give more and more relevant information to the user, the need to integrate new user's needs into the data warehouse process becomes obvious. This challenge arises in applications such as analyzing online store transactions, summarizing dynamic document collections, and profiling Web traffic.

In the following section, we address the problem of personalized OLAP analysis end present our solution to help users to get relevant analyzes from the data warehouse.

## ANALYSIS PERSONALIZATION IN COMPLEX DATA WAREHOUSES

Because of the role of data warehouses in the daily business work of a company, the requirements for design and implementation are often dynamic and subjective. In such a context, it is very helpful to provide users with the most relevant data warehouse model for personalized decision queries. The obtained model must be in continuous evolution with respect to new analysis needs and/or data source changes. In the context of personalized OLAP queries in data warehouses, we thus develop the Rule-based Data Warehouse (R-DW) approach (Bentayeb et al., 2008), in which user aggregation rules help integrate and share user knowledge and allow the warehouse model to evolve by generating new dimension hierarchies. Our approach is based on an extension of the concept of personalization applied to data warehouses, since we deal not only with user preferences but also with new analysis needs. The main consequence of our approach then consists of considering schema evolution in data warehouse models. In this chapter, we particularly focus on XML data warehouse evolution, since XML data warehouses allow for analyzing complex data from the Web.

### Related work: personalization and schema evolution in data warehouses

#### Personalization

Personalization has been mentioned for many years in various domains such as information retrieval and databases. In these domains, personalization usually consists of exploiting user preferences to provide pertinent answers to users. More precisely, in the context of databases, personalization mainly takes the form of adding predicates to queries (Kießling, 2002). In the context of information retrieval, the idea consists of representing user profile with keywords (Domshlak & Joachims, 2007) in order to define a restricted research area. In these domains, personalization then consists of finding pertinent answers within a profusion of data.

Research studies about personalization in data warehouses are recent and constitute an emerging trend. The first work is inspired from the concept of restriction, particularly focused on data visualization and user preference-driven navigation. Bellatrèche et al. (2005) define a dedicated profile that allows refining queries to show only a part of data, which meets the user's preferences. Ravat & Teste (2008) propose a solution to personalize OLAP navigation by exploiting the definition of preferences through weights. In this case, the user assigns weights to the multidimensional concepts to directly get the desired analysis, avoiding a lot of navigation operations. Garrigós et al. (2009) proposed the personalization in data warehouses according to a conceptual point of view with UML.

Through several works exist in data warehouse personalization, we note a lack in integrating user knowledge into warehouse models to take new user analysis needs into account. This constitutes an important issue since only few analysis possibilities are known in the design step of a data warehouse. In this context, we propose to create new analysis axes to meet new user analysis needs.

## Data warehouse model evolution

We can distinguish in the literature two types of approaches that take data warehouse model evolution into account: model updating and temporal modeling. The first approach consists of enriching the data warehouse schema (Blaschka et al., 1999; Hurtado et al., 1999a, 1999b) with adapted evolution operators that allow an evolution of the schema. In this case, only one schema is supported and evolution history is not preserved.

In opposition, the second approach keeps track of schema evolution, by using temporal validity labels. These labels are affixed on dimension instances (Bliujute et al., 1998), on aggregation links (Mendelzon & Vaisman, 2000), or on schema versions (Bebel et al., 2004; Body et al., 2002; Morzy & Wrembel, 2004). In such a data warehouse, each version describes a schema and data at certain periods of time. In order to appropriately analyze multiversion data, an extension to a traditional SQL language is required.

Both approaches do not directly involve users in the data warehouse evolution process, and thus rather constitute a solution for data source evolution than for user analysis need evolution. A personalization process then constitutes a promising research issue to provide relevant analysis for new user needs.

## Data warehouse model evolution for user personalized analysis

In a decision-support context, the system must be user-centered to cope with users' analysis needs. However, in classical decision-support systems based on a data warehouse, user role is limited to data navigation with OLAP. Based on a fixed data warehouse model, navigation allows for answering expected decision queries but offers a limited capacity to take new analysis objectives into account. We propose to personalize the OLAP process to ensure the integration and the exploitation of new analysis needs defined by users.

In this context, we introduce a data warehouse model evolution (i.e. structure and data) in which personalized analysis possibilities can be shared by different users. The personalization process is based on the user knowledge about data aggregation, which is the basis of data organization in multidimensional models. Then, our key idea consists of generating new analysis axes by dynamically creating new dimension hierarchies or extending existing old ones. More precisely, we define new granularity levels inside the data warehouse schema. This provides a real time evolution of dimension hierarchies to cope with personalized analysis needs.

To support our proposed data warehouse model evolution, we define an evolving data warehouse formal model based on aggregation rules, R-DW (Rule-based Data Warehouse), which is independent of any implementation considerations. We first introduced this model in (Bentayeb et al., 2008). Aggregation rules of R-DW allow defining new user needs and consequently creating new granularity levels in the current data warehouse. The R-DW model is composed of a "fixed" part, corresponding to fact tables and the dimensions directly linked to fact tables; and an "evolving" part, defined by aggregation rules. Aggregation rules define aggregation links between two successive levels in a dimension hierarchy and are under the form of "if-then" rules, where an "if" clause defines conditions on the first level and a "then" clause defines the desired (second) level to be created.

We define in the following our global architecture for data warehouse evolution for analysis personalization in which the R-DW model takes place. In this architecture, the user not only navigates within data, but he/she also expresses his/her knowledge to generate new analysis possibilities. Our architecture is composed of four parts corresponding to four phases of our personalized decision support system (Figure 6): (1) users' knowledge acquisition, (2) knowledge integration, (3) data warehouse model update, and (4) OLAP.

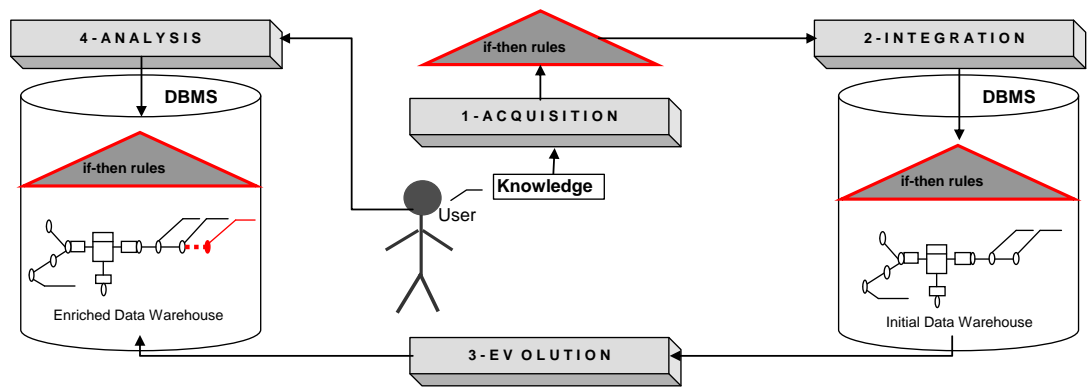

*Figure 6. Architecture for OLAP personalization*

In our approach, we consider a specific user's knowledge, which provides new aggregate data. The acquisition phase aims at collecting user knowledge under the form of "if-then" aggregation rules. Aggregation rules are divided into two types of rules: structure rules and data rules. Structure rules help define the structure of aggregation links, meaning that they allow defining the level to be created, the attributes defining the new level, on what level the new one is based, on what attributes of this lower lever conditions are expressed. Data rules instantiate structure rules, meaning that they define the aggregation link on data themselves. To build a new level, the user defines one structure rule, and various data rules. Data rules define the various instances to be created in the new level and the condition that defines the link with the instances of the lower level. Each data rule corresponds to one instance of the level to be created.

Rules are then integrated within the data warehouse model (integration phase) and used to create new levels in the dimension hierarchy (evolution phase). The last phase, namely on-line analysis, consists of applying decision queries onto the updated data warehouse schema. To validate our data warehouse model evolution, we propose a prototype implemented in the relational context through the WEDriK (data Warehouse Evolution Driven by Knowledge) platform (Favre et al, 2007) under Oracle 10g.

## A user-driven approach for complex data analysis

Since our R-DW model is independent from any implementation, we propose to exploit R-DW to deal with complex data. More precisely, we propose to adapt and apply our approach to complex data warehouses represented in XML format. Thus, we consider the XML data warehouse model presented in a previous section.

Because of the specificities of XML storing, XML data warehouse evolution is not an easy task. Indeed, we have to consider that structure and data are mixed in documents, even if the data warehouse schema is represented in dw-model.xml. Moreover, we have to take the organization of information as a tree into account.

The personalization process consists of creating a new level within a given hierarchy. It implies not only creating this level with required data, but also the links with (an)other level(s). Moreover, we have to deal with both structure and data.

To achieve this evolution in the XML data warehouse, we have to consider the evolution of two documents: the document representing the data warehouse model (dw-model.xml) and the document of the dimension that is concerned by level creation (dimensiond.xml).

The evolution process is based on the use of "if-then" rules defined by the user. For a given structure rule and the set of corresponding data-rules, we are building one granularity level. The interest of the structure-rule is to determine the structure of the level. Thus, we can modify the dw-model.xml document by adding a Level node. The dimensiond.xml document concerned by the evolution of the hierarchy is also modified to include the new level with the various properties and the links with the existing level @Drill-Down if the level is added at the end of a hierarchy, @Drill-Down and @Roll-Up if the new level is inserted between two existing ones.

The implementation requires updates of XML documents. First, a Web interface allows user interaction. The interface helps user define graphically its structure rule and data rules. This interaction is developed using PHP scripts. The evolution of the XML data warehouse requires updates within XML documents that are used to store the data warehouse. Xupdate is a lightweight XML query language for modifying XML data. It is specified by the XML:DBInitiative. It is a simple XML update language used to modify XML content by simply declaring what changes should be made in XML syntax. Various operations can be combined to achieve our evolution, such as inserting an element after, inserting an attribute, updating attribute, etc. Another possibility is to consider the use of DOM (Document Object Model) to achieve the evolution. Indeed, DOM is a platform and language-independent standard object model for representing HTML or XML documents. Thus, it is possible to use PHP scripts with DOM.

As an example, let us consider the study we made with our linguist colleagues. The token frequencies were observed with respect to location in the transcription and speaker sex. The initial data warehouse has been designed according to the CLAPI database and the identified analysis needs. Now, let us suppose that a linguist needs to aggregate frequencies by grouping some locations together. For instance, he wants to know where some tokens appear in the transcription (beginning, middle, and end). Even if this need has not been initially expressed, our personalization process allows for evolving the data warehouse to provide an answer. The linguist is then able to formulate the following aggregation rules (structure rule and the corresponding data rules) through an interface, to express how to aggregate data.

---

Structure rule:

if ConditionOn(location-in-transcription, {location} then Generate(group-of-location, {group-location}

Data rules:     (1)     if location in {'begin', 'end'} then group-location={extreme}

                (2)     if location not in {'begin', 'end'} then group-location={middle}

---

With these rules, the XML data warehouse can be modified: the dw-model.xml document and the dim-time.xml document that corresponds to the dimension that is enriched by a new level. In practice, the dw-model.xml presented in the Figure 4 is modified to include the new level (Figure 7).

```xml
<?xml version="1.0" encoding="utf-8">
<DW-model>
    <dimension id="time-d" path="dim-time.xml">
        <Level id="location-in-transcription">
            <attribute name="location" type="string" />
        </Level>
        <Level id="group-of-location-in-transcription">
            <attribute name="location-group" type="string" />
        </Level>
    </dimension>
    <dimension id="speaker-d" path="dim-speaker.xml">
        <Level id="speaker">
            <attribute name="sex" type="boolean" />
        </Level>
    </dimension>
    <dimension id="transcription-d" path="dim-transcript.xml">
        <Level id="token">
            <attribute name="term" type="string" />
        </Level>
        <Level id="transcription">
            <attribute name="transcription-name" type="string" />
        </Level>
    </dimension>
    <FactDoc id="facts" path="facts.xml">
        <measure id="frequency" type="real" />
        <dimension idref="time-d" />
        <dimension idref="speaker-d" />
        <dimension idref="transcription-d" />
    </FactDoc>
</DW-model>
```

*Figure 7. Updated sample dw-model.xml document*

The dim-time.xml document is also updated to take into account the new level and its instances since the new level concerns the time dimension (Figure 8).

```
<?xml version="1.0" encoding="utf-8">
<dimension dim-id="time-d">
      <Level id="location-in-transcription">
          <Instance id="begin" Roll-up="extreme">
              <attribute id="location" value="begin">
          </Instance>
          <Instance id="middle" Roll-up="middle">
              <attribute id="location" value="middle">
          </Instance>
          <Instance id="end" Roll-up="extreme">
              <attribute id="location" value="end">
          </Instance>
      </Level>
      <Level id="group-of-location-in-transcription">
          <Instance id="extreme" Drill-Down=("begin","end")>
              <attribute id="location-group" value="extreme">
          </Instance>
          <Instance id="middle" Roll-up="middle">
              <attribute id="location-group" value="middle">
          </Instance>
      </Level>
</dimension>
```

*Figure 8. Updated sample dim-time.xml document*

Thus, the linguist is able to compute frequencies with respect to the group of locations and other dimensions, getting an answer to his own analysis needs.

## CONCLUSION

We presented in this chapter a complete solution to warehouse complex data from the Web. Even if much recent research has focused on the design of multidimensional models (Luján-Mora et al., 2006), only few research work addresses the issues of managing and analyzing complex data.

First, we defined a reference XML data warehouse model that unifies and generalizes existing, similar schemas from the literature. This model explicitly takes into account complex features that are possible in XML, but would be intricate to implement in a relational system, such as complex hierarchies and irregular facts.

To provide semantics to OLAP operators, we extend the capabilities of OLAP to description, clustering and explanation by coupling OLAP with data mining techniques. The AHC method is used to define a new OLAP aggregation operator to analyze complex data. To improve navigation into data cubes, we also used the MCA method to reorganize facts inside the data cube by identifying interesting facts. Finally, to help explanation from multidimensional data, we used frequent association rule mining to extract relationships and associations between facts.

To design and build data warehouses, traditional data and goal-driven approaches bear the latent risk of not meeting user requirements. Therefore, user-driven developing approaches seem promising for successful completion of data warehouse projects. In this chapter, we proposed an advanced data warehouse architecture that serves as a modeling framework for user oriented OLAP analyses that take new analysis needs into account. Our main idea is to define a user-oriented evolution of dimension hierarchies in the XML data warehouse. To this end, we proposed a rule-based data warehouse model that presents the advantage of evolving incrementally according to the user's needs.

The perspectives opened by this work are numerous. Let us summarize them. First, we will take inspiration from the principle of $X^3$, Wiwatwattana et al. (2007)'s XML cube operator, to enhance our XOLAP operators and truly make them XML-specific. For instance, we are actually currently working on performing roll-up and drill-down operations onto the ragged hierarchies defined by Beyer et al. (2005). Moreover, to improve personalization and decision query processes, we will also investigate the joint evolution of data sources and analysis needs. Finally, we intend validate our approach experimentally by studying its performance in terms of storage space, response time and algorithms complexity.

## FUTURE RESEARCH ISSUES

The specific characteristics of Web data make the design of BI applications difficult. Such applications help users obtain knowledge from the Web. Content acquisition from the Web can be broken into two phases: information retrieval and information extraction. Web information retrieval is the process of gathering potentially relevant content. Knowledge creation starts from information collected from various sources, then information is normalized and structural and semantic integration must be achieved.

One possible solution to facilitate this task is to extract information from the Web, transform and load it into a Web warehouse, which provides uniform access methods for automatic data processing. Web warehousing is conceptually similar to data warehousing approaches used to integrate relational information from databases. However, the structure of the Web is very dynamic and cannot be controlled by warehouse designers. Web models do not frequently reflect the current state of Web data sources. Thus, Web warehouses must be redesigned at a late stage of development.

From this point of view, two promising research issues can be studied to achieve Web data warehousing:

First, information retrieval and data mining techniques may be used to collect relevant information from the Web and extract knowledge from them. These techniques are not only useful as analysis support tools, but also as modeling support tools. The semantics extracted from Web data can indeed help build relevant multidimensional models and data cubes. Therefore, relevant analyses can be performed and then presented in a form suitable for use by various individuals playing different roles in an organization. Hence, combining information retrieval, data mining techniques and OLAP can help build relevant Web data cubes.

Second, semantic Web technologies may be employed to collect semantically appropriate data from the Web. More specifically, the Resource Description Framework (RDF) may be used for describing Web data, and the Web Ontology Language (OWL) for creating an ontology that restricts the semantics of RDF descriptions. We think that, using this approach, we can implement a piece of software that assists users in designing a suitable OLAP schema and performing data extraction, transformation and loading (ETL).